**Ratiometric Organic Fibers for Localized and Reversible Ion Sensing with Micrometer-Scale Spatial Resolution**

*Loretta L. del Mercato\*, Maria Moffa, Rosaria Rinaldi, Dario Pisignano\**

Dr. Loretta L. del Mercato, CNR NANOTEC - Istituto di Nanotecnologia, Polo di Nanotecnologia c/o Campus Ecotekne, via Monteroni 73100 Lecce, Italy.
E-mail: loretta.delmercato@nanotec.cnr.it
Dr. Maria Moffa, Prof. Dr. Rosaria Rinaldi, Prof. Dr. Dario Pisignano
Istituto Nanoscienze-CNR, Euromediterranean Center for Nanomaterial Modelling and Technology (ECMT), via Arnesano 73100, Lecce, Italy
Prof. Dr. Rosaria Rinaldi, Prof. Dr. Dario Pisignano
Dipartimento di Matematica e Fisica "Ennio De Giorgi", Università del Salento, Via Arnesano, 73100 Lecce, Italy

E-mail: loretta.delmercato@nanotec.cnr.it
E-mail: dario.pisignano@unisalento.it

**Abstract**

A fundamental issue in biomedical and environmental sciences is the development of sensitive and robust sensors able to probe the analyte of interest, under physiological and pathological conditions or in environmental samples, and with very high spatial resolution. In this work, novel hybrid organic fibers that can effectively report the analyte concentration within the local microenvironment are reported. The nanostructured and flexible wires are prepared by embedding fluorescent pH sensors based on seminaphtho-rhodafluor-1-dextran conjugate. By adjusting capsule/polymer ratio and spinning conditions, the diameter of the fibers and the alignment of the reporting capsules are both tuned. The hybrid wires display excellent stability, high sensitivity, as well as reversible response, and their operation relies on effective diffusional kinetic coupling of the sensing regions and the embedding polymer matrix. These devices are believed to be a powerful new sensing platform for clinical diagnostics, bioassays and environmental monitoring.

# 1. Introduction

pH plays a crucial role in many biological processes including wound healing, tissue regeneration, and cancer. For instance, acidity is found at the surface of skin[1,2] and in inflammatory sites,[3] and is also associated with bone resorption.[4,5] There is substantial evidence showing that an acidic microenvironment regulates cellular phenotype and, notably, acidic extracellular pH (pHe) is a major feature of tumor tissues.[6] Therefore, pH is an important control parameter for the maintenance of cellular viability and for improving tissue functions. The development of sensitive and selective nanostructured probes for proton ions ($H^+$) on both microscopic (i.e., with high spatial resolution) and macroscopic (i.e., over areas of several $cm^2$) scales is consequently highly desirable for the analysis and monitoring of a number of physiological and pathological conditions.[7,8] Similarly, pH sensors are largely demanded for the fast and accurate monitoring of water quality.[9]

Flexible nanowires are nowadays of great scientific and industrial interest because of their potential application in different fields, including photonics and electronics,[10] catalysis and electrocatalysis,[11,12] energy,[13] and biomedicine.[14–16] In this framework, electrospinning is a straightforward, fast, and low-cost method to fabricate fibers which are extremely long and which show diameters ranging from microns to tens of nm, with tailored composition. Owing to their size and surface features, the resulting fibers might exhibit high surface-to-volume ratios and porosity. These properties, favoring the transport of small molecules and fluids across fibrous scaffolds, make such fibers especially attractive for the development of ultrasensitive sensors.[17–21] Recently, oppositely charged polyelectrolytes have been used for obtaining pH-responsive fibers relying on reversible swelling-deswelling mechanisms.[22] In addition, numerous techniques have been also suggested for spinning nanofibers with enhanced porosity and surface area, involving the use of low boiling point solvents,[23–25] humid environments,[26] or post-treatment of blends or composites by selective removal of one of the components.[27] Importantly, electrified jets can conveniently carry sensing micro- and nanoparticles, which following dispersion in polymer solutions can be spun to define stable functional regions in continuously delivered filaments. Various colloids, including microencapsulated aqueous reservoirs,[28,29] microporous organic/inorganic, biomolecular[30] and polymeric[31] particles, have been incorporated in electrified jets.[32] In particular, light-emitting capsule dispersions allow multicompartment nanostructured materials to be generated with novel functionalities which are strictly dictated by the particle properties.

Here, we present the development of fluorescent, pH-sensing organic fibers usable as biomaterial scaffold for microenvironment analysis. By monitoring fluorescence changes of specific fiber regions, ratiometric measurements of changes in local proton concentration can be performed in a rapid and non-invasive manner, and with high spatial control. Capsule-based sensors have previously been developed by loading various ion-sensitive probes,[33–40] and related barcoded sensors have been reported for the simultaneous measurement of $H^+$, $Na^+$, and $K^+$ concentration in test tubes.[41] Most studies validated the use of these sensors for the analysis of intracellular pH (pHi) changes.[34,36,38,40,42–44] In contrast, here we focus on designing a mechanically robust and flexible architecture suited for pHe analyses. The unique combination of a permeable fiber matrix, sensitivity to pH, and ratiometric and reversible response gives these structures a great potential for various applications. In particular, they could find use as flexible and versatile tools for quantitatively monitoring the changes in $H^+$ concentration that impact three-dimensional biological response.[45] The realized hybrid fibers are based on stably incorporated capsules and are prepared through a multistep procedure combining electrospinning with layer-by-layer (LbL) capsule assembly. LbL assembly allows for the formation of multilayer films and capsules through the alternating adsorption of polyanions and polycations.[46] The composition and properties of the resulting structures can be tailored by using polymers with different functions, as well as by stimuli-responsive nanomaterials (i.e., silver, gold and magnetic nanoparticles).[47–49] The presented material fully preserves the response of pristine sensors to local pH changes. Capsules are orderly packed within the wires, and the functional scaffolds can be successfully fabricated on a large scale (~$10^2$ cm$^2$) for biological and environmental analysis.

## 2. Results and Discussion

### 2.1. Fabrication of pH-sensing hybrid organic wires

We fabricated pH-sensing hybrid organic wires with different diameter and appreciable surface porosity by a highly volatile DCM/ACE mixture, trapping capsule-based fluorescent sensors within PLLA. The fabrication procedure is illustrated in **Scheme 1**. The first step involves the conjugation of a pH-indicating fluorescent probe to a macromolecule (Scheme 1a).

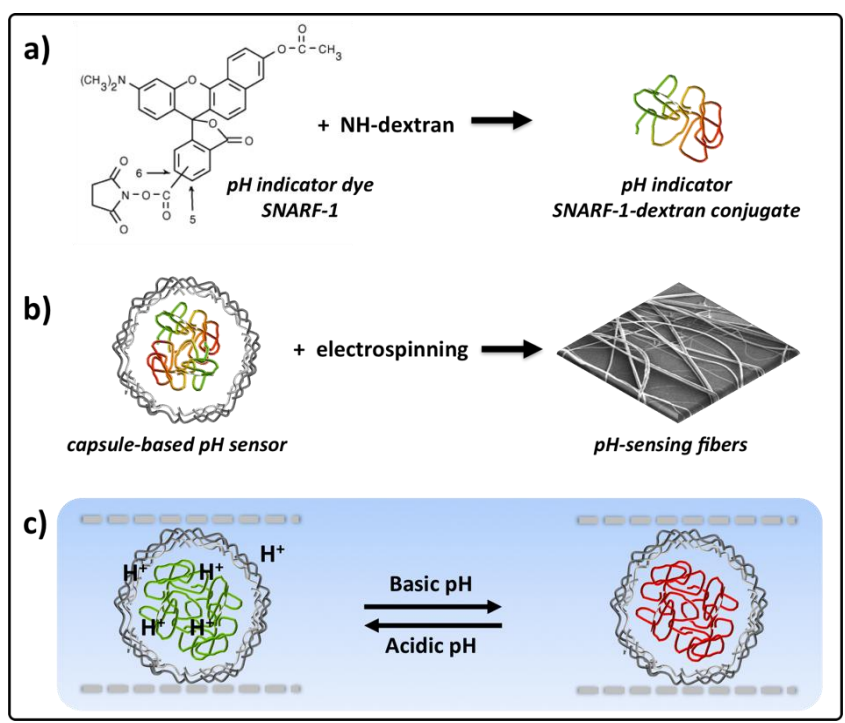

**Scheme 1.** Fabrication of pH-sensing wires. a) Sketch showing the conjugation of the ratiometric fluorescent pH indicator dye SNARF-1 to nonfluorescent aminodextran. b) Schematic representation of a capsule-based pH sensor (carrying SNARF-1-dextran conjugate in the cavity) and SEM image showing the wires obtained via electrospinning. c) Schematic view of the response of a pH-sensing hybrid wire to the local environment. Objects are not drawn to scale.

We selected the fluorescent pH indicator SNARF-1, validated for *in vitro* and *in vivo* sensing,[50–56] and covalently linked it to aminodextran. SNARF-1 emits at two different wavelengths (~ 580 nm and 640 nm).[57] Its response to local pH is determined by the ratio of the fluorescence intensities at 580 nm and 640 nm while exciting the dye at one wavelength (between 488 nm and 530 nm). The protonated form emits in the yellow-orange region (540-580 nm), whereas deep red emission (620-640 nm) is observed from the basic form.[58] The second step involves the LbL assembly of capsule-based pH sensors[33,34] (Scheme 1b).

Capsules which contain SNARF-1-dextran conjugate in their cavities and a multilayer shell of three bilayers [(PSS/PAH)$_3$] were produced with average diameter of (4.2 ± 0.1) μm. The pH-sensitivity was assessed by recording the capsule fluorescent response to various pHs from 4-9 (Figure S1 in the Supporting Information). The third step yields hybrid fibers (Scheme 1b), using PLLA as template polymer because of its processability, biocompatibility, and good mechanical properties.[59–61] For calibration sensing experiments, the capsule dispersion was mixed with the PLLA DCM/ACE solution and then spun onto glass cover slides (24×60 mm$^2$) (Figure S2). The concentration of capsules and PLLA solutions, the solvents as well as the spinning conditions (applied voltage=8 kV, needle-collector distance=10 cm, flow rate=0.5 mL/h) were adjusted to obtain controlled diameter and prevent capsule aggregation, thus leading to a specific surface topology as specified below. In the resulting hybrid wires, the capsule-based pH-sensors cannot leak out of the polymer filaments. Ions, such as protons (H$^+$), can pass through the wire outer regions[18,62] and the multilayer capsule shells,[63–66] and thereby get sensed by the fluorescent pH indicator (Scheme 1c). In other words, ions (-OH$^-$ and H$^+$) are allowed to protonate and deprotonate the embedded pH indicator SNARF-1-dextran conjugate. As the red-to-green ratio of the fluorescence signal from cavities depends on the local pH around each capsule, fibers at acidic pH will display capsules with green fluorescence (false color), whereas fibers at basic pH will display capsules with predominantly red fluorescence (Scheme 1c).

## 2.2. Hybrid fiber morphology

**Figure 1a** shows a photograph of free-standing fibers bearing pH sensors. The scanning electron microscopy (SEM) micrographs in Figures 1b and 1d display wires with different diameters. In all cases, spinning PLLA from DCM/ACE solutions led to fibers with a rough and porous surface structure. For fibers with average pristine diameter up to about 300 nm, which are obtained by a 8% (w/w) PLLA DCM/ACE solution (Figure 1b), the individual filaments are quite smooth, whereas a local increase of surface roughness can be appreciated in the regions corresponding to pH-sensing features (Figures 1c and S3).

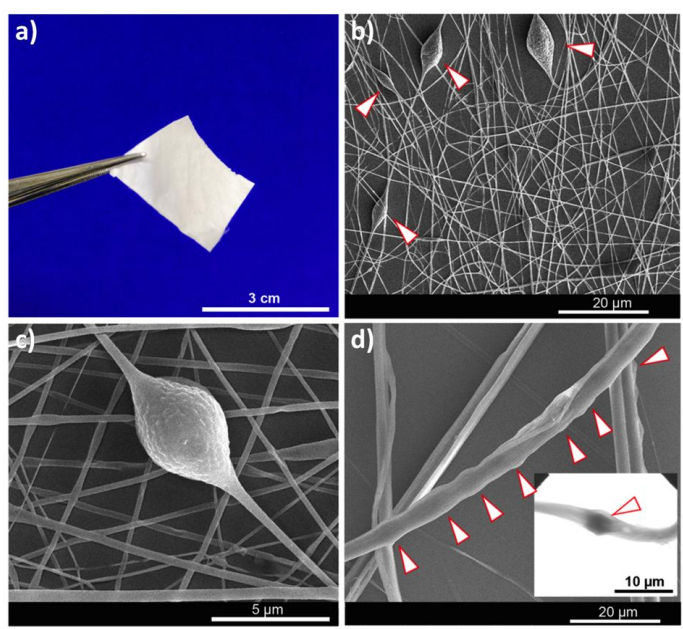

**Figure 1.** Morphology of pH-sensing hybrid organic wires. a) Photograph of a free-standing sample of fibers. (b-d) SEM micrographs. b) Random non-woven mat of fibers. c) High-magnification SEM micrograph of an individual fiber carrying an embedded capsule. The fiber surface roughness, increasing in the capsule region, can be appreciated. d) Zoomed STEM micrograph. Arrowheads indicate capsule-based pH sensors. Inset: individual capsule.

In addition, in the same regions small pores are found by SEM. This morphology can be induced by the locally enhanced solvent evaporation, due to the thinner polymer skin, and to the associated outer shell collapse and buckling phenomena.[67,68] Notably, capsules appear to retain a spherical shape (Figure S3 in the Supporting Information) suggesting the internal retention of water even following wire solidification. Elongated features are instead found on the body of fibers with diameter of 3-4 μm, in segments nearby sensing regions (Figures 1d and S3b). These fibers were realized by mixing a 12% (w/w) PLLA DCM/ACE solution, and they carry a higher amount of capsules aligned along their longitudinal axis (Figure 1d, arrowheads). This particular morphology promoted by emulsion electrospinning can be very useful for optical sensing in microfluidics, favoring the flow of liquids along the elongated nanostructures similarly to directional water collection in wet-rebuilt spider silk,[69] and presenting localized pores enhancing analyte diffusion. By scanning transmission electron microscope (STEM), the position of each capsule can be precisely identified in individual wires due to the high contrast between the multilayer $(PSS/PAH)_3$ shell (darker region in the inset of Figure 1d) and PLLA (bright regions). Finally, CLSM images reveal spherical objects of average size (4.0±0.1) μm (**Figure 2**), with no appreciable structural damage or deformation.

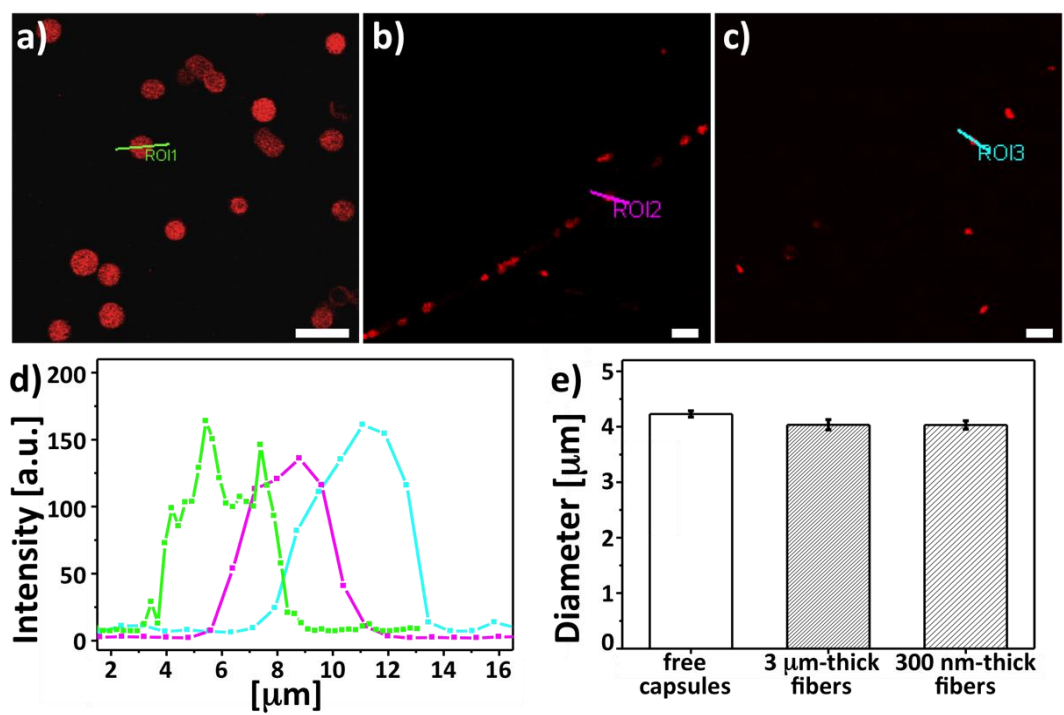

**Figure 2.** Confocal analysis of the sensing regions. The CLSM micrographs show a) fluorescent free capsules, capsules in b) 3 µm-thick fibers, and in c) 300 nm-thick fibers, respectively. Scale bar = 10 µm. d) Intensity profiles of the region of interest (ROIs 1-3) marked with a line in the CLSM micrographs reported in (a-c). e) Average diameter of free capsules (4.2±0.1 µm), capsules in 3 µm-thick (4.0±0.1 µm), and in 300 nm-thick fibers (4.0±0.1 µm), respectively. Columns represent mean ± standard error of mean (number of capsules analysed = 50).

## 2.3. pH-sensitivity of hybrid organic wires

In order to calibrate the response to pH changes, the wires were exposed to buffer solutions with known pHs, from 4 to 9 (**Figure 3a** and Figures S4 and S5). Figure 3a shows CLSM images of three distinct regions of 3 µm-thick fibers with aligned fluorescent pH-sensors exposed to acidic (5), neutral (7) and basic (9) pHs. Following contact with pH-adjusted buffer solution, the fluorescence signals in the yellow (shown as false color in green) and red channels were sequentially recorded.

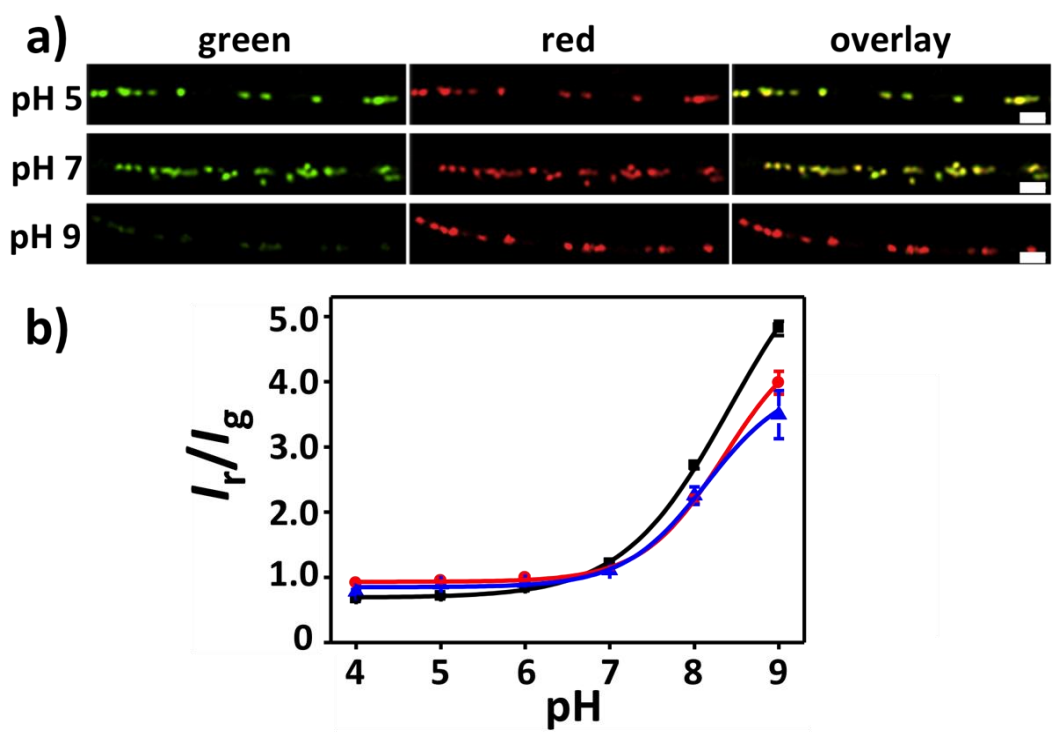

**Figure 3.** pH-sensitivity of hybrid organic fibers. a) Fluorescent micrographs of wires exposed to acidic (5), neutral (7) and basic (9) pHs, imaged upon casting a 10 μL aliquot of buffer. The individual green (false color, 540-610 nm) and red (620-700 nm) channels of CLSM micrographs are shown, followed by overlay of the two channels. Scale bars = 10 μm. b) Typical response curves to pH changes. Free capsule-based pH sensors (black squares), sensors 300 nm-thick (red circles) and in 3 μm-thick wires (blue triangles), respectively. The red-to-green ratio (false colors) of the fluorescence signal $I_r/I_g$ is here plotted *vs.* the pH of the solution. The data points correspond to the mean ± standard error of mean, calculated over at least 35 capsules. Data were fitted with a sigmoidal function,[33] having points of inflection at pH = 8.41 (black fit), 8.35 (red fit) and 8.14 (blue fit), respectively. CLSM micrographs were collected as described in "Experimental Section".

The emission clearly depends on the local pH, namely, PLLA-wrapped capsules predominantly emit in the green (false color) at acidic pH and in the red region of the visible spectrum at basic pH, respectively, in accordance with the photophysical properties of the indicator dye.[58] Emission variations occur on a timescale of seconds, indicative of the diffusive kinetics. By measuring, for each fluorescence image, the ratio of red ($I_r$: 620–700 nm) to yellow ($I_g$: 540–610 nm, shown in green) emission signal, the response curves [$I_r/I_g$(pH)] of the pristine capsules and of the capsule-based fibers were obtained (Figure 3b). The limited sensitivity found at pH values below 5.5 is not surprising in view of the known features of the used pH indicator,[58] and it can be largely improved by using indicator dyes alternative to SNARF-1[58] (Figure S6). Importantly, loaded capsules show good sensitivity especially in the near neutral or alkaline pH region (i.e., pH range from 6 to 9). No significant changes in intensity ratios can be observed for fibers with different diameters, the main differences being instead related to the proton diffusive rates and to the degree of fatigue exhibited following many pH sensing cycles.

**Figure 4a** shows the fluorescent micrographs and the $I_r/I_g$-values when the pH is switched from 9 and 4 for three consecutive cycles. Before adding the new solutions, the exposed area is rinsed with Milli-Q water. Notably, wires show an excellent robustness to both the pH switches and the rinsing steps. From the fluorescent images, it can also be seen that sensing elements remain stably entrapped in the wires without undergoing any change in morphology and position (Figure 4a). Fibers show a good reversibility to the pH switches (Figure 4b). After three cycles, the $I_r/I_g$-read out allows us to clearly discriminate between pH 9 and pH 4. Notably, by collecting the fluorescence response of distinct areas with low integration times ($\cong$ 50 μs) after leaving samples in dark for a few minutes at pH 9, thus avoiding photobleaching, no remarkable fatigue effect, namely no significant variation in the corresponding $I_r/I_g$ values, is found in the sensors. Overall, the good reproducibility of the ratiometric sensing suggests a reliable ionic diffusion through both the PLLA polymer matrix and the polyelectrolyte multilayer shell of the capsules as well as robust encapsulation. Data reported in Figure S7 support these conclusions, and also highlight that the response to pH 4 is very fast, whereas the response to pH 9 proceeds more slowly (3-5 min, Figure 4c). The fluorescence changes are in agreement with a diffusional uptake according to the Equation (1):

$$C(t) = C_\infty (Dt/\pi d^2)^{0.5}, \qquad (1)$$

where $C$ indicates the amount of ions inducing the fluorescence changes, $t$ is time, $D$ is an effective diffusion coefficient, $d$ is the fiber radius, and $C_\infty$ is a constant (continuous line in Figure 4c).

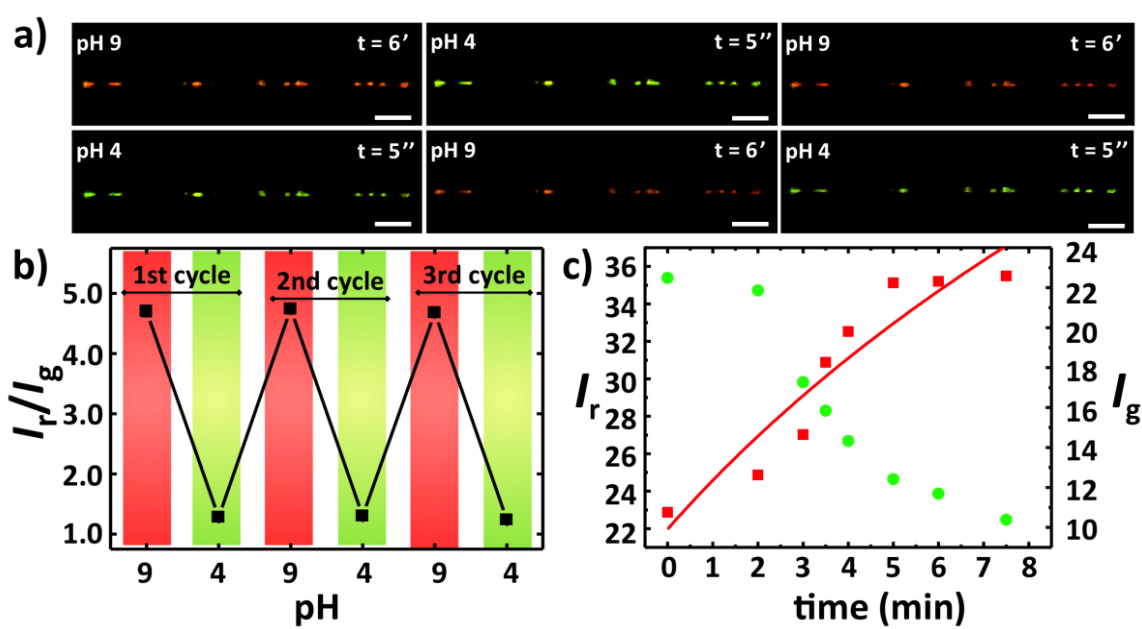

**Figure 4.** Reversibility and time response to pH switches. a) Typical reversibility response of hybrid fibers, imaged via CLSM after addition of a drop of solution at pH 9. Next, a drop of solution at pH 4 was deposited onto the same region. The cycle was repeated up to 20 times. Overlay of green (false color, 540-610 nm) and red (620-700 nm) channels recorded at each tested pH. The $t$ values for each frame indicate the time interval after the application of a pH change. Average fiber diameter = 3 µm. Scale bars = 20 µm. b) Red-to-green ratio (false colors) of the fluorescence signal $I_r/I_g$ *vs.* the pH of the solution. c) Fiber temporal response to changes from low (4) to high (9) pH. Left axis for the fluorescent signals measured for red (620-700 nm, squares) channel and right axis for the green (false color, 540-610 nm, circles) channel. Average fiber diameter = 3 µm. Continuous line: best fit to red signal channel data by Equation (1).

Furthermore, the remarkably asymmetric behavior in the temporal response observed upon switching fibers from high to low pH and vice versa unveils a complex underlying mechanism. On one side, previous studies have evidenced the influence of multilayer shells on the diffusion times of ions into capsules.[63,64,70] Indeed, multilayer shells act as semi-permeable barriers between the entrapped analyte-sensitive probe and the external medium, which may increase the response time of the fluorescent reporter by affecting the ion diffusion. In our case, capsules with PAH as outermost layer were employed. PAH chains are highly protonated at low pH, whereas at high pH the ammonium groups are more deprotonated. As a result, at pH 9 the multilayer shells adopt a loopy conformation that may partially decrease the permeation of $H^+$ ions through the capsule shell. The resulting time response would be faster upon exposure to acidic pHs. However, the difference in the measured time response in this case would be of the order of $10^{-1}$ s,[64] i.e. much lower than that found here (i.e. seconds *vs.* minutes). Consequently, diffusional aspects involving the fiber polymer matrix might have a significant role in the observed switching behavior, which could be made asymmetric upon either reducing or increasing the environment pH due to various concomitant mechanisms. These effects can include local swelling of the polymer matrix promoted by the loopy shells of the capsules at high pH, which would then lead to a fast diffusional behavior when pH is subsequently decreased. Also, local hydrolysis of ester bonds of PLLA is possible at alkaline conditions, which, together with the plasticizer effect of water,[71] would reduce the stability of polymer chains in the fibers and locally enhance molecular mobilities. With respect to that, water pinning on fibers and the related reduced spreading of water, observed in our system, might significantly contribute to increasing the ultimate stability, decreasing the interaction of the polymer matrix with the aqueous environment.[72] Overall, the operation and stability of our wires are found to be based on the interplay between the polyelectrolyte multilayers and the embedding polymer matrix. In fact, this conclusion is supported by the response time of capsules embedded in thinner (~300 nm) fibers, which are rapidly activated by both pHs, 4 and 9, thus highlighting the reduced role of ionic diffusion through the polymer matrix compared to thicker fibers (Figure S8). Here, the larger surface to volume ratio provides a generally smaller distance for $H^+$ diffusion within the PLLA matrix and through the multilayer shells to reach fluorescent reporters.

Taken together, the above results suggest that these systems may have widespread applications in pH sensing devices. For instance, they could be used for measuring, *in-situ* and with μm-spatial resolution, the extracellular proton concentration in complex heterogeneous constructs and cultured three-dimensional scaffolds, unveiling the effects of drug treatments or the release of acidic by-products from individual cell response. While this kind of analysis would allow fully understanding of the role of $H^+$, and of other relevant analytes ($O_2$, $Ca^{2+}$, $Cu^{2+}$), several ion-sensitive fluorescent indicators show interference from analytes different from their desired targets, thus potentially leading to erroneous optical outputs. For all these systems, ultimate parallel sensing of multiple analytes by simultaneously monitoring multiple fluorescence lifetimes,[73,74] as would be promptly implemented through these fiber architectures, is of significant interest. Other applications can be found in optical monitoring of pH for assessing water quality, and for measuring the concentration of ions in fluids carried in microfluidic channels, since these fibers and membranes made of them can be easily integrated with microfluidic devices and architectures. Furthermore, the controlled delivery of therapeutics could be integrated within these systems, by co-loading fibers with drugs and stimuli-responsive capsules. Limitations are set by the temporal response of the hybrid fibers, which may become critical for sensing fast pH changes, and by the stability of the realized constructs under flow conditions where significant shear stress is present.

## 3. Conclusion

In conclusion, we demonstrated a novel ratiometric fiber material that can be used for localized and reversible pH sensing. Sensing elements are stably aligned within fibers, which show good sensitivity in the pH range 4-9, with the most sensitive dynamic pH range from 6 to 9. The wire diameter was found to influence the time response to pH changes according to an effective diffusional kinetic interplay with capsule external shells. This platform represents a valuable model for the real-time detection of localized $H^+$ gradients, and of other relevant ions and small molecules, in the fields of biological and environmental research. The hybrid wires can be tailored by simply changing the probe reporters with a wide range of fluorescent indicators,[33,36,37,40,41,75,76] moreover multiplex read-out can be achieved by embedding capsules sensitive to different analytes (e.g., $H^+$, $O_2$, glucose, etc.) into individual fibers. In particular, the co-loading of analyte sensors and stimuli-responsive drug-loaded capsules is considered a key next step toward the use of these composites for medical-related applications including tissue engineering and wound healing.

## 4. Experimental Section

*Chemicals*: Poly(sodium 4-styrenesulfonate) (PSS, ~70.000 MW), poly(allylamine hydrochloride) (PAH, ~56.000 MW), calcium chloride dehydrate (CaCl2, 147.01 MW), sodium carbonate (Na$_2$CO$_3$, 105.99 MW), ethylenediaminetetraacetic acid disodium salt dehydrate (EDTA), polylactic acid (PLLA, 85.000–160.000 MW), glutaraldehyde solution grade I (50% in H2O), sodium cacodylate trihydrate, dichloromethane (DCM) and acetone (ACE) were purchased from Sigma-Aldrich. Aminodextran (500.000 MW) and 5-(and-6-)-carboxy-seminaphtho-rhodafluor-1 acetoxymethyl ester acetate (SNARF-1, 567.5508 MW) were obtained from Invitrogen. The pH sensing measurements were performed by using commercial standard citric acid/sodium hydroxide buffer solutions of different pH values from 4 to 9 (Fluka, Sigma). All chemicals were used as received. Ultrapure water with a resistance greater than 18.2 M$\Omega$ cm was used for all experiments.

*Synthesis of Capsule-Based pH-Sensors*: The pH indicator dye SNARF-1 was conjugated to the nonfluorescent aminodextran and subsequently loaded inside porous calcium carbonate ($CaCO_3$) microparticles obtained via co-precipitation of $Na_2CO_3$ (0.33 M) and $CaCl_2$ (0.33 M) solutions. The as-produced SNARF-1-dextran filled $CaCO_3$ particles were then coated by multiple layer-by-layer (LbL) assembly of the oppositely charged PSS (2 mg mL$^{-1}$, 0.5 M NaCl, pH = 6.5) and PAH (2 mg mL$^{-1}$, 0.5 M NaCl, pH = 6.5) polyelectrolytes. This procedure was repeated until 6 layers were assembled around the spherical microparticles, thus providing a multilayer shell (PSS/PAH)$_3$. In the last step, the sacrificial $CaCO_3$ cores were removed by complexation with EDTA buffer (0.2 M, pH 7). Finally, the SNARF-1-dextran filled multilayer polyelectrolyte capsules were stored as suspension in 2 mL of Milli-Q water at 4°C. After core removal the number of capsules per volume was determined by direct counting in a defined volume with a hemocytometer chamber under an inverted optical microscope. From one synthesis we obtained $9.47 \times 10^8$ capsules/mL.

*Fabrication of pH-sensing hybrid organic wires*: Fibers were prepared by electrospinning using DCM/ACE (80:20 v/v), dissolving PLLA at room temperature with overnight stirring. Capsules were mixed with the solution (100 µL of suspension per 1 mL) followed by thorough mixing, transfer into a 1 mL syringe and delivery to the tip of a 27 gauge stainless steel needle by a syringe pump (Harvard Apparatus, Holliston, MA) with a feeding rate of 0.5 mL/h. A positive high-voltage of 8 kV was applied to the solution. Fibers were collected on glass cover slides positioned on a grounded target (Al foil) at a distance of 10 cm from the extruding needle.

*Morphological Characterization of hybrid organic wires*: Before SEM analysis, fibers underwent chemical fixation and drying. Specifically, samples were fixed in 2.5% glutaraldehyde buffer for 30 min, and then rinsed twice in cacodylate buffer solution. The wires were then dehydrated in increasing concentrations of ethanol, transferred to an increasing graded series of hexamethyl–disilazane and sputtered-coated with a nano-chromium film. The morphology was finally evaluated by SEM (FEI Company, Hillsbora, Oregon, USA). The average diameter of the fibers was calculated from the SEM images using an imaging software (Image J), analyzing a total number of at least 100 fibers. To calibrate the capsule-based pH sensors and to assess their potential use in probing local pH, small volume of capsules suspensions in Milli-Q water (5 µL) were added to pH-adjusted buffers (30 µL). Final pHs were 4, 5, 6, 7, 8, and 9. Samples were mixed, allowed to equilibrate for 10 min, and transferred to a conventional glass slide for subsequent CLSM analysis.

*pH Sensing Assays*: To study the fluorescence response of PLLA fibers to different pH solutions, an aliquot of pH-adjusted buffers (10 µL) was deposited above a defined region of the fibers and the exposed area was instantly imaged via CLSM. To study the reversibility of the fluorescence response to switches of pH, an aliquot (10 µL) of the buffer at pH 9 was deposited above a defined region of the hybrid fiber. The threated area was instantly scanned via CLSM. Subsequently, the drop was aspirated away by means of a micropipette and the exposed area was washed repeatedly by Milli-Q water before adding an equal volume (10 µL) of the buffer at pH 4. The same area was then scanned again via CLSM. The pH switch cycles (from 9 to 4) were repeated up to 20 times for 3 µm-thick fibers and up to 40 times for 300 nm-thick fibers. In case of 3 µm-thick fibers, the area exposed to pH 9 was scanned multiple times (every minute) in order to monitor the fluorescence response of the hybrid fibers over the time (up to 6 minutes). In Figure S7, four time points at pH 9 (second and third cycles of pH switch) are reported for sake of clarity. The acquired images were then analyzed as described below.

All fluorescence microscopy images were collected using a Leica confocal laser scanning microscope (CLSM) (TCS SP5; Leica, Microsystem GmbH, Mannheim, Germany). Thicker and thinner fibers were observed with a 63×/1.40 and a 40×/1.25 oil-immersion objectives, respectively. The pH indicator SNARF-1 was excited by using the 514-nm line of an argon ion laser (50%), and its emission was recorded between 540 and 610 nm ("green" channel, false color) and 620 and 700 nm ("red" channel). To avoid cross talk between the green and red detection channels, the images were collected sequentially in the $x$, $y$, $z$ planes (scan speed 400 Hz, pinhole aperture set to 1 Airy, integration time $\cong$ 50 µs).

*Image Analysis*: The emissions of pH-sensor capsules at different pH values, before and after embedment within PLLA fibers, were evaluated with Image J software. First, a region of interest (ROI) of the same shape and size, was selected in the centre of the individual capsules.[33] Then, the ratio of red ($I_r$; 620-700 nm) to yellow ($I_g$; 540-610 nm, shown in green) fluorescence $I_r/I_g$(pH) was calculated according to Equation (2):

$$\frac{I_r}{I_g}(pH) = \frac{(I_r - I_{r,bkgd})}{(I_g - I_{g,bkgd})},$$
(2)

Finally, the obtained red-to-green ratio (false colors) of the fluorescence signal $I_r/I_g$ from the capsule cavities were plotted versus the known pHs for determining the response curves for each analyzed system. Data points were fitted with a Boltzmann sigmoidal function.[33] A total number of 50 ROIs (free capsules) and 35 ROIs (capsules embedded in the fibers) were analyzed for each pH value to calculate the average red/green ratio at a specific ionic concentration. All fluorescence ratios were corrected by subtracting the background (bkgd) fluorescence from each image (red and green channels) according to equation 1. As a control, untreated PLLA fibers were imaged, under different pHs, to ensure that the eventual polymer autofluorescence would not interfere with pH quantification (see Figure S9). In all reported images the fluorescence is displayed in false colours.

**Acknowledgements**

The research leading to these results has received funding from the European Research Council under the European Union's Seventh Framework Programme (FP/2007-2013)/ERC Grant Agreement n. 306357 (ERC Starting Grant "NANO-JETS"). The Apulia Network of Public Research Laboratories Wafitech (9) and the National Operational Programme for Research and Competitiveness (PONREC) 'RINOVATIS' (PON02_00563_3448479) are also acknowledged. M. Ferraro and R. Manco are acknowledged for initial experimental trials.

[1]    H. Lambers, S. Piessens, A. Bloem, H. Pronk, P. Finkel, *Int J Cosmet Sci* **2006**, *28*, 359.

[2]    S. Dikstein, A. Zlotogorski, *Acta Derm. Venereol.* **1994**, 18.

[3]    A. Lardner, *J. Leukoc. Biol.* **2001**, *69*, 522.

[4]    N. S. Krieger, K. K. Frick, D. A. Bushinsky, *Curr. Opin. Nephrol. Hypertens.* **2004**, *13*, 423.

[5]    D. A. Bushinsky, K. K. Frick, *Curr. Opin. Nephrol. Hypertens* **2000**, *9*, 369.

[6]    Y. Kato, S. Ozawa, C. Miyamoto, Y. Maehata, A. Suzuki, T. Maeda, Y. Baba, *Cancer Cell Int.* **2013**, *13*, 89.

[7]    A. Jonitz, K. Lochner, T. Lindner, D. Hansmann, A. Marrot, R. Bader, *J. Mater. Sci. Mater. Med.* **2011**, *22*, 2089.

[8]    B. van der Sanden, M. Dhobb, F. Berger, D. Wion, *J. Cel.l Biochem.* **2010**, *111*, 801.

[9]    E. M. V. Hoek, A. K. Ghosh, *Nanotechnology Applications for Clean Water*, **2009**.

[10]  P. Wang, Y. P. Wang, L. M. Tong, *Light Sci. Appl.* **2013**, *2*, 1.

[11]  A. C. Patel, S. X. Li, C. Wang, W. J. Zhang, Y. Wei, *Chem. Mater.* **2007**, *19*, 1231.

[12]  X. X. Yan, L. Gan, Y. C. Lin, L. Bai, T. Wang, X. Q. Wang, J. Luo, J. Zhu, *Small* **2014**, *10*, 4072.

[13]  V. Thavasi, G. Singh, S. Ramakrishna, *Energy Environ. Sci.* **2008**, *1*, 205.

[14]  N. Bhattarai, Z. S. Li, J. Gunn, M. Leung, A. Cooper, D. Edmondson, O. Veiseh, M. H. Chen, Y. Zhang, R. G. Ellenbogen, M. Q. Zhang, *Adv. Mater.* **2009**, *21*, 2792.

[15]  L. Moroni, R. Schotel, D. Hamann, J. R. De Wijn, C. A. Van Blitterswijk, *Adv. Funct. Mater.* **2008**, *18*, 53.

[16]  Y. Ji, K. Ghosh, B. Li, J. C. Sokolov, R. A. F. Clark, M. H. Rafailovich, *Macromol. Biosci.* **2006**, *6*, 811.

[17]  B. Ding, M. Wang, J. Yu, G. Sun, *Sensors (Basel)* **2009**, *9*, 1609.

[18]  X. Y. Wang, C. Drew, S. H. Lee, K. J. Senecal, J. Kumar, L. A. Sarnuelson, *Nano Lett.* **2002**, *2*, 1273.

[19]  J. Yoon, S. K. Chae, J. M. Kim, *J. Am. Chem. Soc.* **2007**, *129*, 3038.

[20]  I. D. Kim, A. Rothschild, *Polym. Adv. Technol.* **2011**, *22*, 318.

[21]  N. G. Cho, H.-S. Woo, J.-H. Lee, I.-D. Kim, *Chem. Commun.* **2011**, *47*, 11300.

[22]  M. Boas, A. Gradys, G. Vasilyev, M. Burman, E. Zussman, *Soft Matter* **2015**, *11*, 1739.

[23]  Y. Miyauchi, B. Ding, S. Shiratori, *Nanotechnology* **2006**, *17*, 5151.

[24]  M. Bognitzki, W. Czado, T. Frese, A. Schaper, M. Hellwig, M. Steinhart, A. Greiner, J. H. Wendorff, *Adv. Mater.* **2001**, *13*, 70.

[25]  S. Megelski, J. S. Stephens, D. B. Chase, J. F. Rabolt, *Macromolecules* **2002**, *35*, 8456.

[26]  C. L. Casper, J. S. Stephens, N. G. Tassi, D. B. Chase, J. F. Rabolt, *Macromolecules* **2004**, *37*, 573.

[27]  Y. You, J. H. Youk, S. W. Lee, B. M. Min, S. J. Lee, W. H. Park, *Mater. Lett.* **2006**, *60*, 757.

[28]  I. G. Loscertales, A. Barrero, I. Guerrero, R. Cortijo, M. Marquez, A. M. Ganan-Calvo, *Science* **2002**, *295*, 1695.

[29]  E. H. Sanders, R. Kloefkorn, G. L. Bowlin, D. G. Simpson, G. E. Wnek, *Macromolecules* **2003**, *36*, 3803.

[30]   B. Dong, M. E. Smith, G. E. Wnek, *Small* **2009**, *5*, 1508.

[31]   E. Jo, S. W. Lee, K. T. Kim, Y. S. Won, H. S. Kim, E. C. Cho, U. Jeong, *Adv. Mater.* **2009**, *21*, 968.

[32]   D. Crespy, K. Friedemann, A. M. Popa, *Macromol. Rapid Commun.* **2012**, *33*, 1978.

[33]   L. L. del Mercato, A. Z. Abbasi, W. J. Parak, *Small* **2011**, *7*, 351.

[34]   O. Kreft, A. M. Javier, G. B. Sukhorukov, W. J. Parak, *J. Mater. Chem.* **2007**, *17*, 4471.

[35]   L. I. Kazakova, L. I. Shabarchina, S. Anastasova, A. M. Pavlov, P. Vadgama, A. G. Skirtach, G. B. Sukhorukov, *Anal. Bioanal. Chem.* **2013**, *405*, 1559.

[36]   X. X. Song, H. B. Li, W. J. Tong, C. Y. Gao, *J. Colloid Interface Sci.* **2014**, *416*, 252.

[37]   L. I. Kazakova, L. I. Shabarchina, G. B. Sukhorukov, *Phys. Chem. Chem. Phys.* **2011**, *13*, 11110.

[38]   K. Liang, S. T. Gunawan, J. J. Richardson, G. K. Such, J. W. Cui, F. Caruso, *Adv. Healthc. Mater.* **2014**, *3*, 1551.

[39]   J. Q. Brown, M. J. McShane, *Ieee Eng. Med. Biol. Mag.* **2003**, *22*, 118.

[40]   P. Zhang, X. X. Song, W. J. Tong, C. Y. Gao, *Macromol. Biosci.* **2014**, *14*, 1495.

[41]   L. L. del Mercato, A. Z. Abbasi, M. Ochs, W. J. Parak, *ACS Nano* **2011**, *5*, 9668.

[42]   P. Rivera-Gil, S. De Koker, B. G. De Geest, W. J. Parak, *Nano Lett.* **2009**, *9*, 4398.

[43]   R. Hartmann, M. Weidenbach, M. Neubauer, A. Fery, W. J. Parak, *Angew. Chemie-International Ed.* **2015**, *54*, 1365.

[44]   M. De Luca, M. M. Ferraro, R. Hartmann, P. Rivera-Gil, A. Klingl, M. Nazarenus, A. Ramirez, W. J. Parak, C. Bucci, R. Rinaldi, L. L. del Mercato, *ACS Appl. Mater. Interfaces* **2015**, *15*, 15052.

[45]   M. A. Schwartz, C. S. Chen, *Science* **2013**, *339*, 402.

[46]   G. Decher, *Science* **1997**, *277*, 1232.

[47]   L. L. del Mercato, E. Gonzalez, A. Z. Abbasi, W. J. Parak, V. Puntes, *J. Mater. Chem.* **2011**, *21*, 11468.

[48]   L. L. del Mercato, M. M. Ferraro, F. Baldassarre, S. Mancarella, V. Greco, R. Rinaldi, S. Leporatti, *Adv. Colloid. Interface Sci.* **2014**, *207*, 139.

[49]   P. Rivera Gil, L. L. del Mercato, P. del-Pino, A. Munoz-Javier, W. J. Parak, *Nano Today* **2008**, *3*, 12.

[50]   Y. Wu, W. Zhang, J. Li, Y. Zhang, *Am. J. Nucl. Med. Mol. Imaging* **2013**, *3*, 1.

[51]   J. C. Tseng, H. A. Benink, M. G. McDougall, I. Chico-Calero, A. L. Kung, *Curr. Chem. Genomics* **2012**, *6*, 48.

[52]   R. C. Hunter, T. J. Beveridge, *Appl. Env. Microbiol.* **2005**, *71*, 2501.

[53]   S. Schlafer, J. E. Garcia, M. Greve, M. K. Raarup, B. Nyvad, I. Dige, *Appl. Env. Microbiol.* **2015**, *81*, 1267.

[54]   I. Bartsch, E. Willbold, B. Rosenhahn, F. Witte, *Acta Biomater.* **2014**, *10*, 34.

[55]   S. Tanaka, S. Chu, M. Hirokawa, M. H. Montrose, J. D. Kaunitz, *Gut* **2003**, *52*, 775.

[56]   V. Estrella, T. A. Chen, M. Lloyd, J. Wojtkowiak, H. H. Cornnell, A. Ibrahim-Hashim, K. Bailey, Y. Balagurunathan, J. M. Rothberg, B. F. Sloane, J. Johnson, R. A. Gatenby, R. J. Gillies, *Cancer Res.* **2013**, *73*, 1524.

[57]   J. E. Whitaker, R. P. Haugland, F. G. Prendergast, *Anal. Biochem.* **1991**, *194*, 330.

[58]   J. Y. Han, K. Burgess, *Chem. Rev.* **2010**, *110*, 2709.

[59]   F. Yang, R. Murugan, S. Wang, S. Ramakrishna, *Biomaterials* **2005**, *26*, 2603.

[60]   H. G. Zhu, J. Ji, M. A. Barbosa, J. C. Shen, *J. Biomed. Mater. Res. Part B-Applied Biomater.* **2004**, *71B*, 159.

[61]   L. J. Zhang, T. J. Webster, *Nano Today* **2009**, *4*, 66.

[62]   M. L. Wang, G. W. Meng, Q. Huang, Y. W. Qian, *Environ. Sci. Technol.* **2012**, *46*, 367.

[63]   G. B. Sukhorukov, M. Brumen, E. Donath, H. Mohwald, *J. Phys. Chem. B* **1999**, *103*, 6434.

[64]   S. Carregal-Romero, P. Rinklin, S. Schulze, M. Schafer, A. Ott, D. Huhn, X. Yu, B. Wolfrum, K. M. Weitzel, W. J. Parak, *Macromol. Rapid Commun.* **2013**, *34*, 1820.

[65]   A. A. Antipov, G. B. Sukhorukov, *Adv. Colloid. Interface Sci.* **2004**, *111*, 49.

[66]   Q. Y. Tang, A. R. Denton, *Phys. Chem. Chem. Phys.* **2015**, *17*, 11070.

[67]   S. Koombhongse, W. X. Liu, D. H. Reneker, *J. Polym. Sci. Part B-Polymer Phys.* **2001**, *39*, 2598.

[68]   C. L. Pai, M. C. Boyce, G. C. Rutledge, *Macromolecules* **2009**, *42*, 2102.

[69]   Y. M. Zheng, H. Bai, Z. B. Huang, X. L. Tian, F. Q. Nie, Y. Zhao, J. Zhai, L. Jiang, *Nature* **2010**, *463*, 640.

[70]   Q. Tang, A. R. Denton, *Phys. Chem. Chem. Phys.* **2014**, *16*, 20924.

[71]   H. Tsuji, K. Sumida, *J. Appl. Polym. Sci.* **2001**, *79*, 1582.

[72]   M. B. Ma, W. L. Zhou, *Ind. Eng. Chem. Res.* **2015**, *54*, 2599.

[73]   A. Z. Abbasi, F. Amin, T. Niebling, S. Friede, M. Ochs, S. Carregal-Romero, J. M. Montenegro, P. R. Gil, W. Heimbrodt, W. J. Parak, *ACS Nano* **2011**, *5*, 21.

[74]   K. Kantner, S. Ashraf, S. Carregal-Romero, C. Carrillo-Carrion, M. Collot, P. del Pino, W. Heimbrodt, D. J. De Aberasturi, U. Kaiser, L. I. Kazakova, M. Lelle, N. M. de Baroja, J. M. Montenegro, M. Nazarenus, B. Pelaz, K. Peneva, P. R. Gil, N. Sabir, L. M. Schneider, L. I. Shabarchina, G. B. Sukhorukov, M. Vazquez, F. Yang, W. J. Parak, *Small* **2015**, *11*, 896.

[75]   R. Srivastava, R. D. Jayant, A. Chaudhary, M. J. McShane, *J. Diabetes Sci. Technol.* **2011**, *5*, 76.

[76]   Y. Lvov, A. A. Antipov, A. Mamedov, H. Mohwald, G. B. Sukhorukov, *Nano Lett.* **2001**, *1*, 125.

# Supporting Information

**Ratiometric Organic Fibers for Localized and Reversible Ion Sensing with Micrometer-Scale Spatial Resolution**

*Loretta L. del Mercato\*, Maria Moffa, Rosaria Rinaldi, Dario Pisignano\**

Dr. Loretta L. del Mercato, CNR NANOTEC - Istituto di Nanotecnologia, Polo di Nanotecnologia c/o Campus Ecotekne, via Monteroni 73100 Lecce, Italy.
E-mail: loretta.delmercato@nanotec.cnr.it
Dr. Maria Moffa, Prof. Dr. Rosaria Rinaldi, Prof. Dr. Dario Pisignano
Istituto Nanoscienze-CNR, Euromediterranean Center for Nanomaterial Modelling and Technology (ECMT), via Arnesano 73100, Lecce, Italy
Prof. Dr. Rosaria Rinaldi, Prof. Dr. Dario Pisignano
Dipartimento di Matematica e Fisica "Ennio De Giorgi", Università del Salento, Via Arnesano, 73100 Lecce, Italy

E-mail: loretta.delmercato@nanotec.cnr.it
E-mail: dario.pisignano@unisalento.it

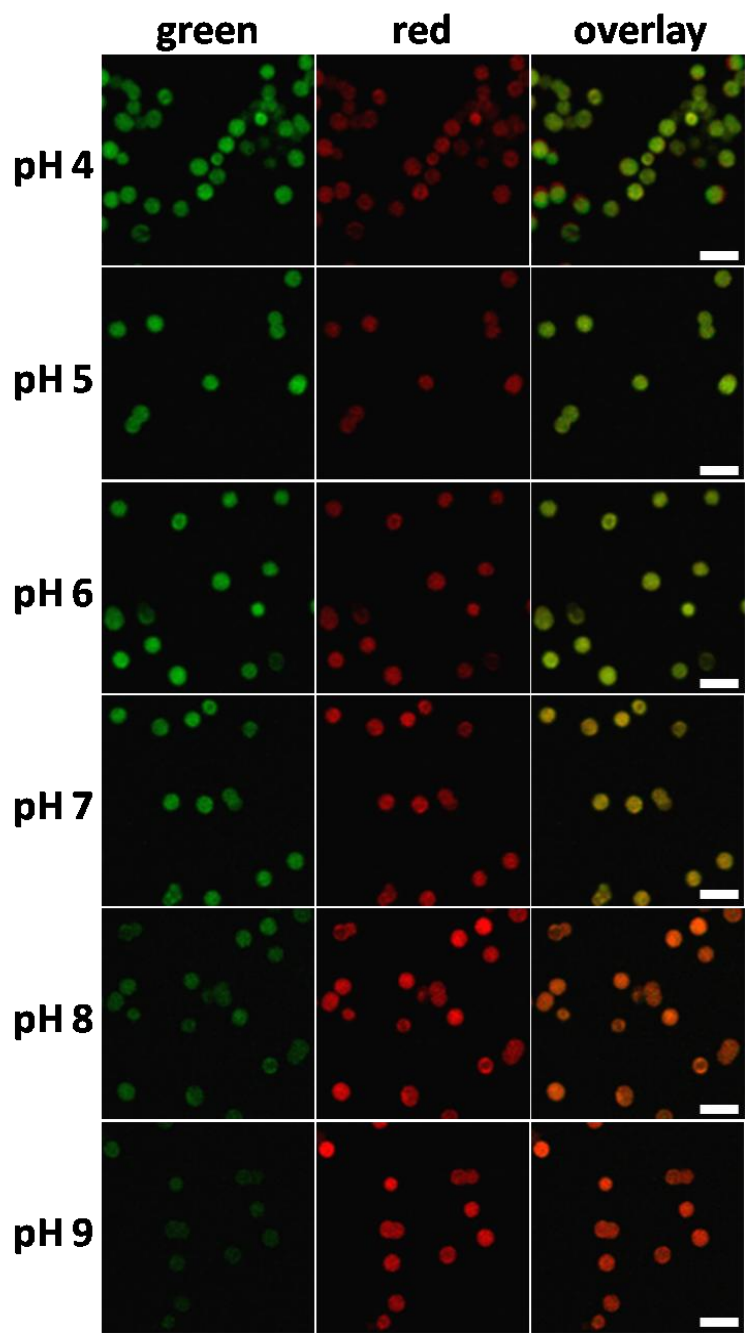

**Figure S1.** Fluorescent micrographs of capsule-based pH sensors at different pHs. The individual green (false color, 540-610 nm) and red (620-700 nm) are shown, followed by overlay of the two channels. Scale bars = 5 μm.

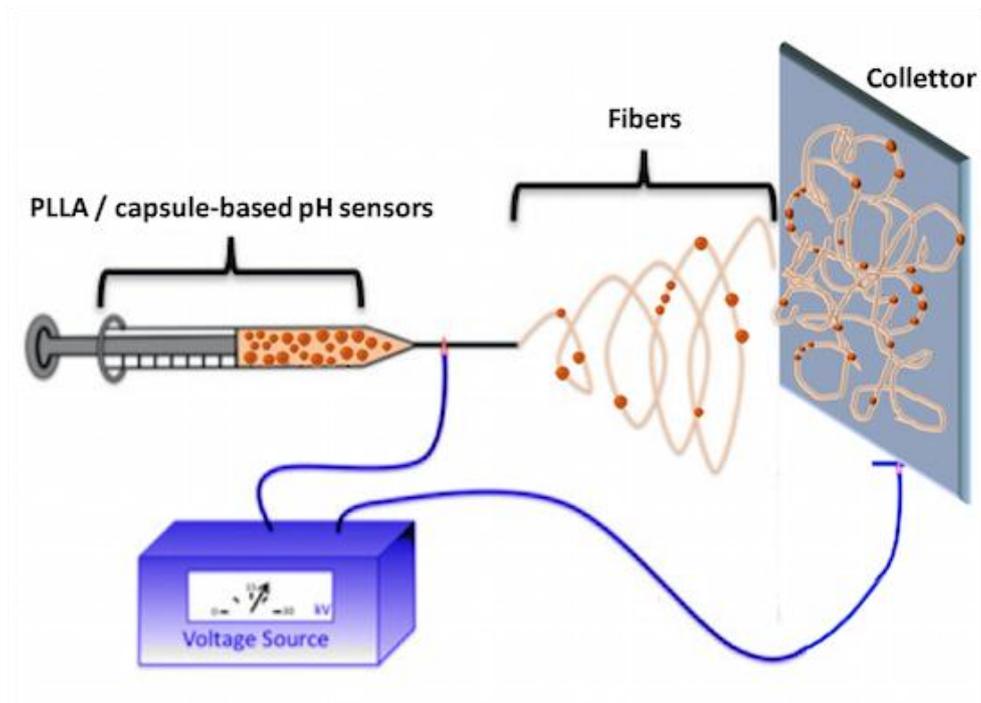

**Figure S2.** Schematic drawing of the electrospinning setup.

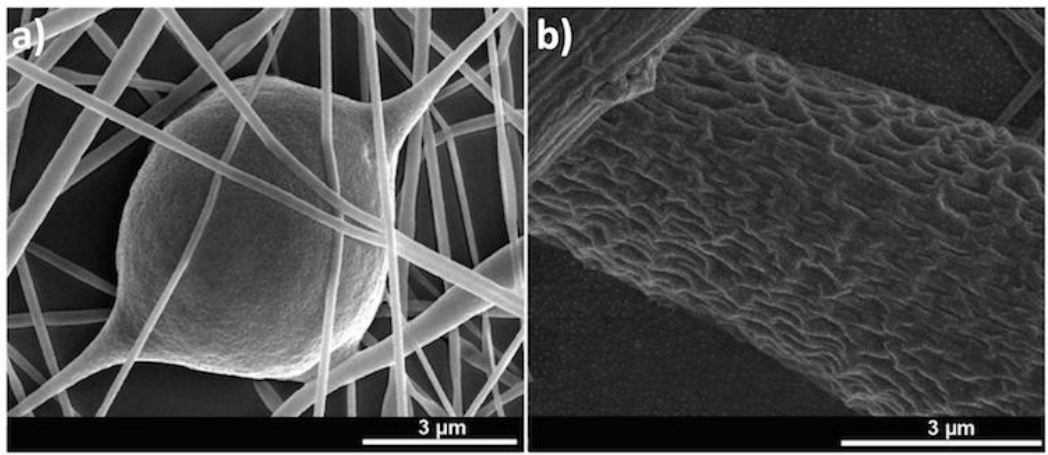

**Figure S3.** SEM micrographs showing the fiber porosity in the sensing region (a) and the roughness on the body of individual wires (b).

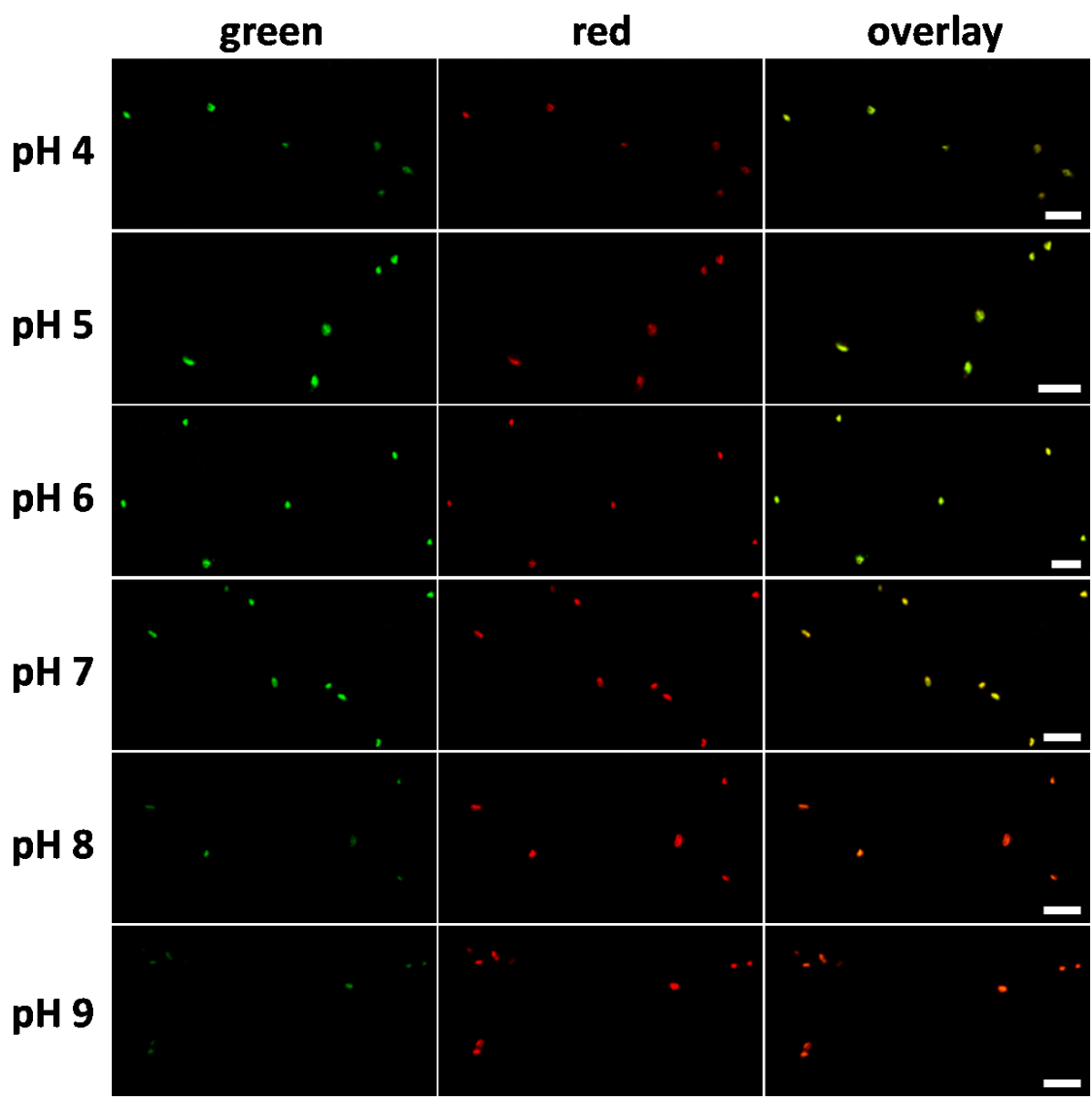

**Figure S4.** Fiber fluorescence micrographs. Average fiber diameter = 300 nm. The individual green (false color, 540-610 nm) and red (620-700 nm) channels together with channel overlay are shown. Scale bars = 20 μm.

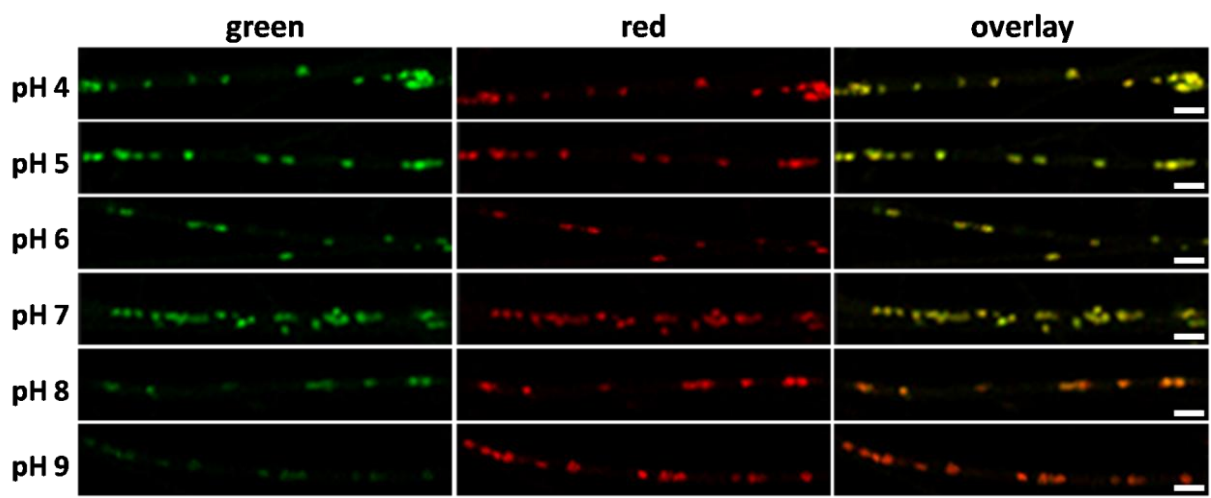

**Figure S5.** Fiber fluorescence micrographs. Average fiber diameter = 3 µm. The individual green (false color, 540-610 nm) and red (620-700 nm) together with channel overlay are shown. Scale bars = 10 µm.

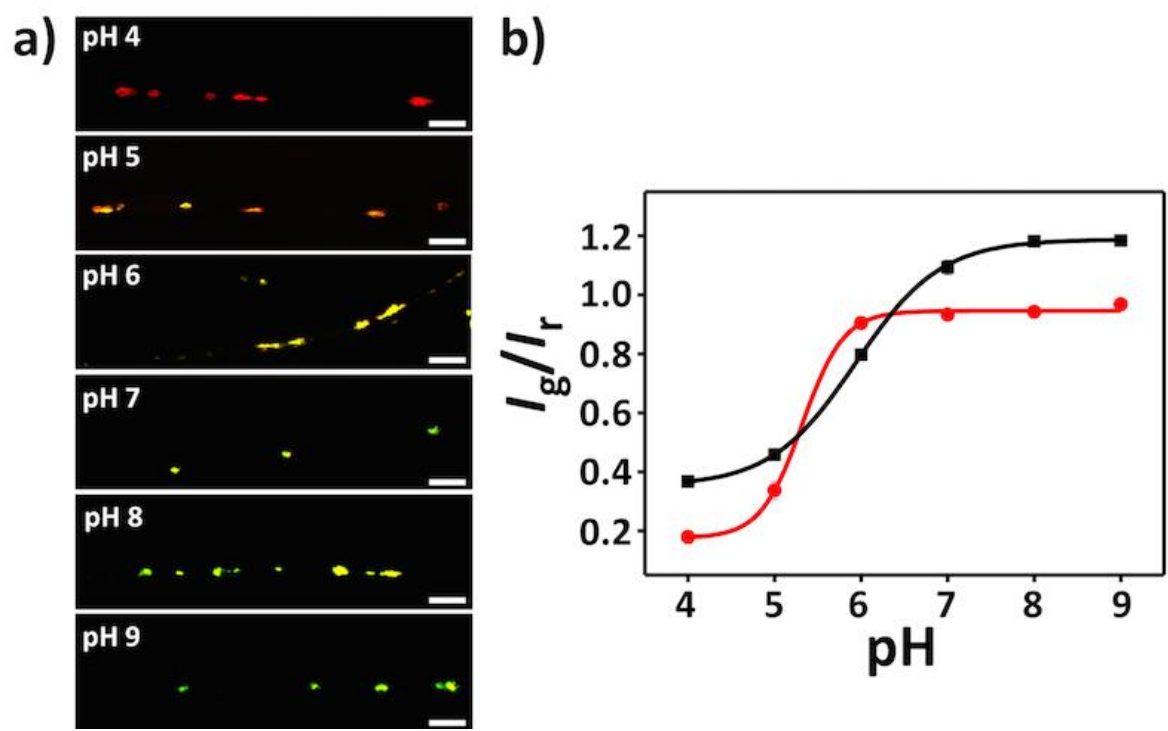

**Figure S6**. Hybrid fibers embedding capsules loaded with fluorescein 5(6)-isothiocyanate (FITC)- and rhodamine B isothiocyanate (RITC)-dextran conjugates. The pH-sensitive dye, FITC (Mw = 389.38 Da, Sigma), and the reference dye, RITC (Mw = 536.08 Da, Sigma), were covalently linked to the nonfluorescent aminodextran and subsequently co-loaded into the cavities of the capsules as previously described.[1,2] Next, the capsules were embedded within 3 μm-thick PLLA fibers, according to the procedure described in the "Experimental Section" for SNARF-1-dextran conjugate loaded capsules. Finally, the pH-sensitivity of the free capsules and of the fibers was monitored by recording the fluorescent response to various pHs from 4 to 9. (a) CLSM micrographs showing the pH-dependence of fluorescence in pH-adjusted cell medium ($\lambda_{exc}$ FITC = 488 nm, $\lambda_{exc}$ RITC = 543 nm). Overlays of the fluorescence channels (green channel: 505−530 nm, red channel: >560 nm) are reported. By increasing the pH of the medium the overall color of the pH-sensing capsules shifts from "red" over "yellow" to "green" as result of the pH dependence of FITC (indicator dye) and the insensitivity of RITC (reference dye). Scale bars: 10 μm. (b) Ratiometric calibration curve of free capsule-based sensors (black squares) and sensors in 3 μm-thick wires (red circles), by fluorescence intensity ratio of green and red channels derived from CLSM micrographs. The green-to-red ratio (false colors) of the fluorescence signal $I_g/I_r$ is here plotted *vs.* the pH of the solution. The data points correspond to the mean ± standard error of mean, calculated over at least 35 capsules. Data were fitted with a sigmoidal function,[2] having points of inflection at pH = 5.94 (black fit) and 5.32 (red fit), respectively.

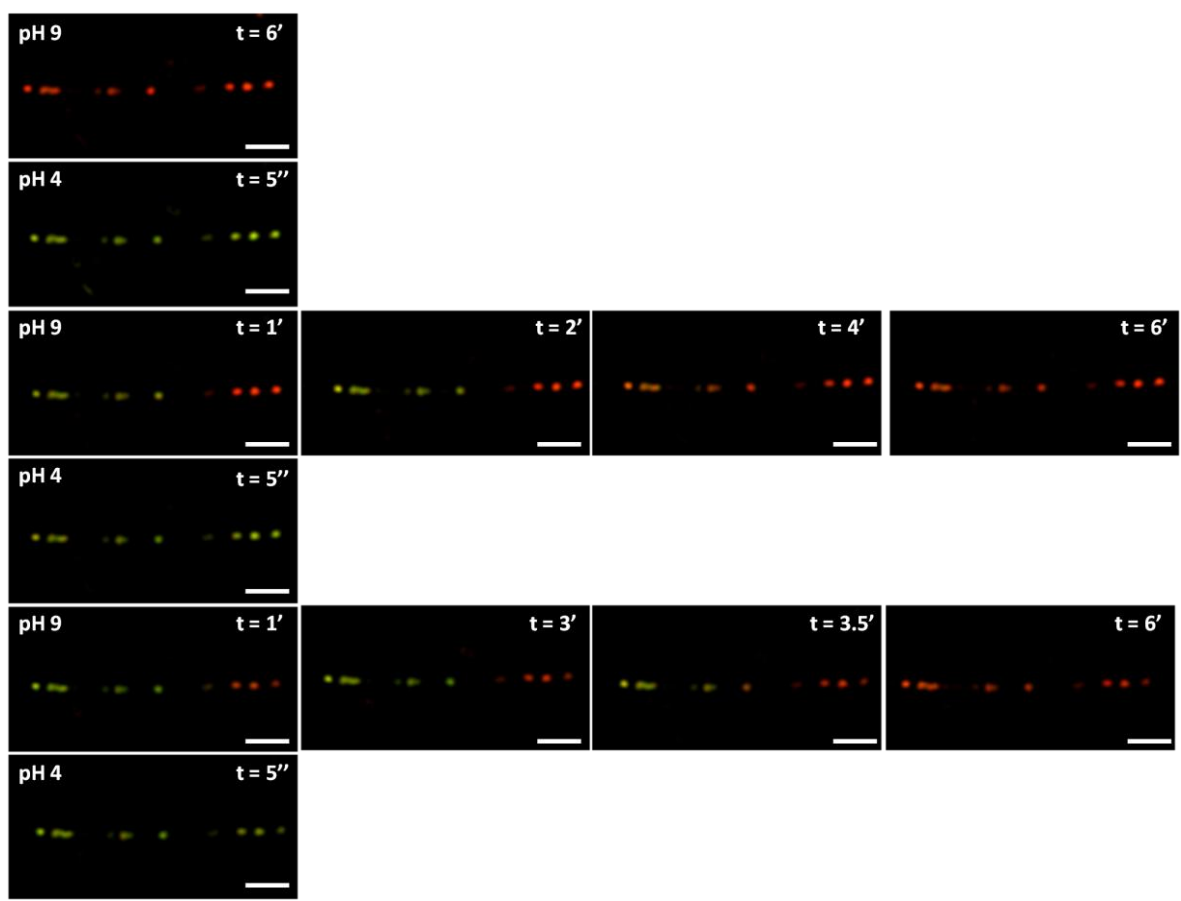

**Figure S7.** Fluorescence time response to pH variations. The CLSM micrographs show the overlays of the green (false color, 540-610 nm) and red (620-700 nm) channels. The *t* values for each frame indicate the time after the application of a pH change. The analysis show a slower response of 3 μm-thick fibers to basic pHs (9) compared to acidic pHs (4). Average fiber diameter = 3 μm. Scale bars = 20 μm.

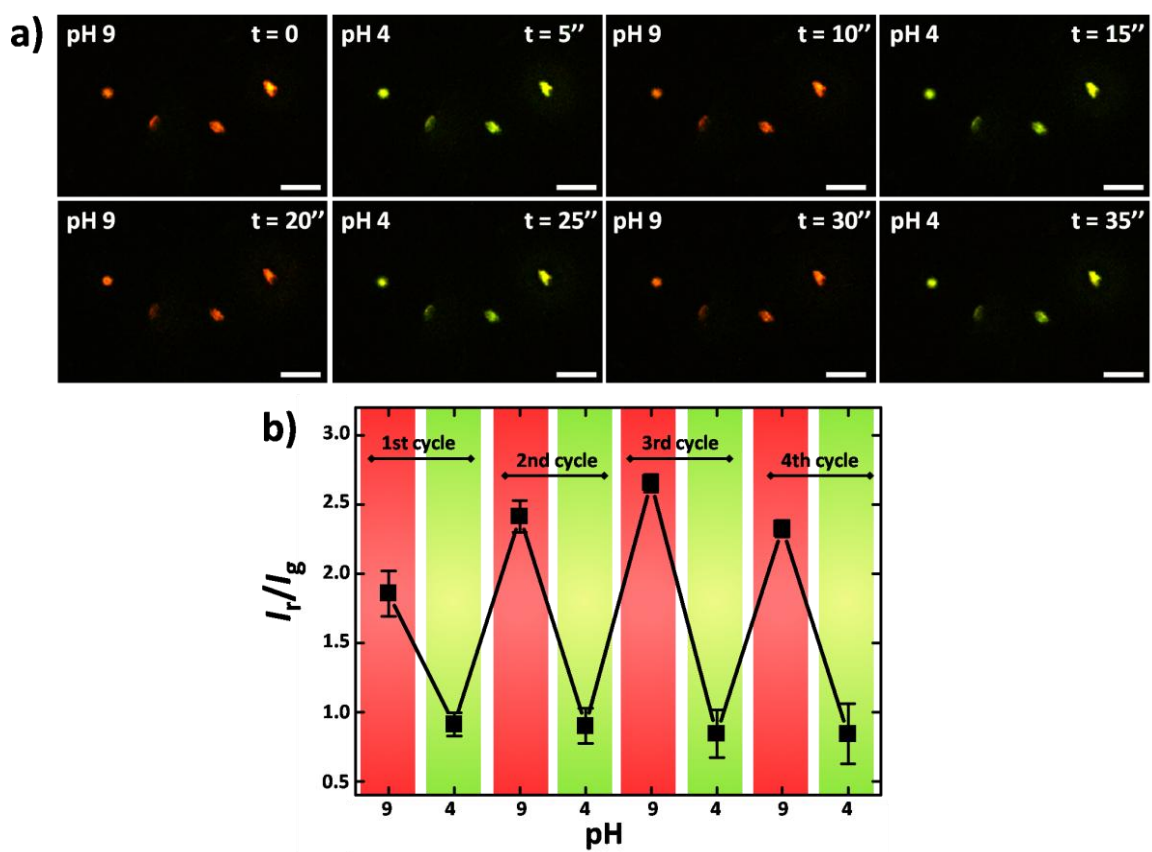

**Figure S8.** Reversible response of pH-sensing. Fibers were imaged via CLSM after addition of a drop of solution at pH 9. Next, a drop of solution at pH 4 was deposited onto the same region. The cycle was repeated up to 40 times. (a) Overlay of green (false color, 540-610 nm) and red (620-700 nm) channels recorded at each tested pH. The $t$ values for each frame indicate the time after the application of a pH change. Average fiber diameter = 300 nm. Scale bars = 10 µm. (b) Red-to-green ratio (false colors) of the fluorescence signal $I_r/I_g$ *vs.* pH.

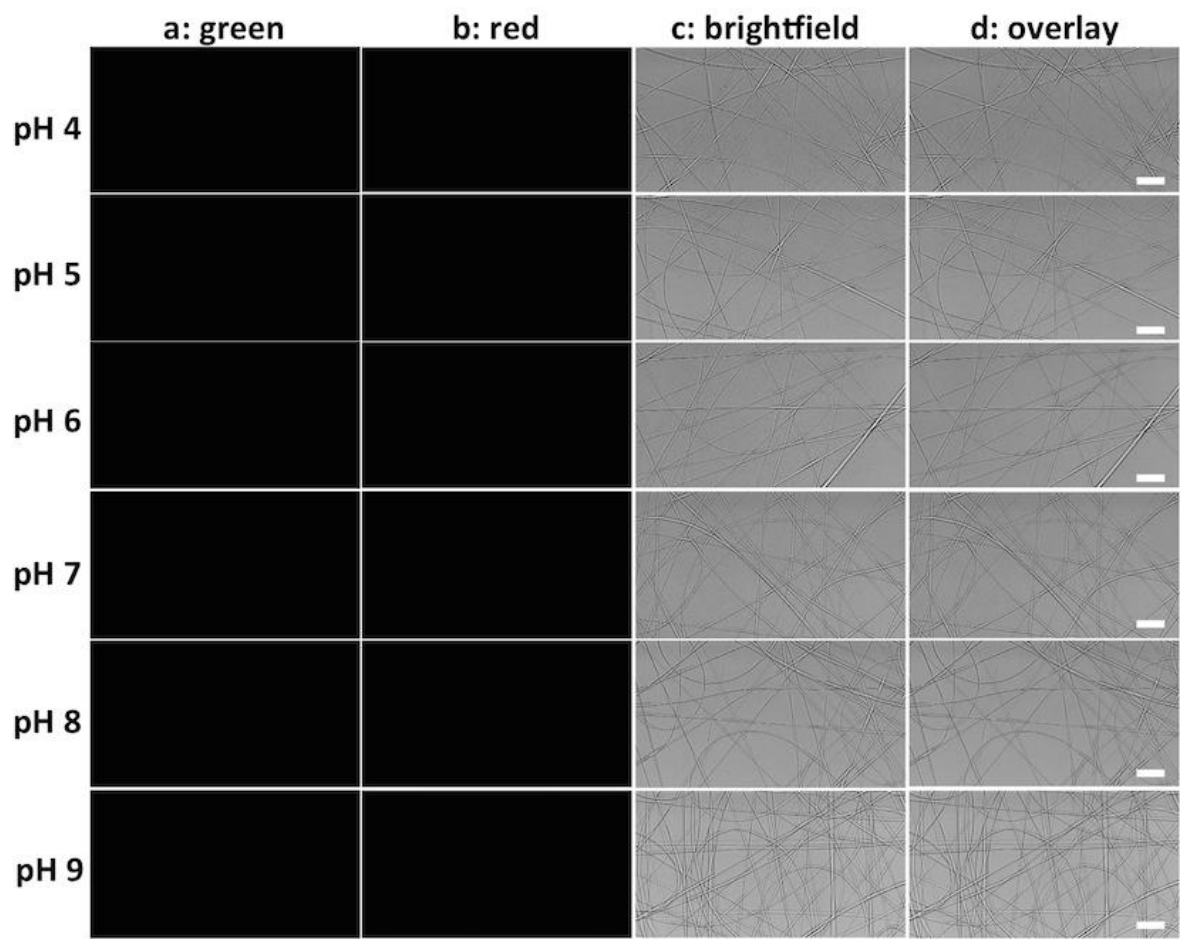

**Figure S9.** Control experiments. CLSM images of PLLA fibers under different pHs. 20 µL droplets of pH-adjusted buffers (4, 5, 6, 7, 8, and 9) were deposited onto a region of 3 µm-thick fibers (without capsules) for 20 minutes; emission was collected in green channel ($\lambda_{em}$ = 540–610 nm, a) and red channel ($\lambda_{em}$ = 620-700 nm, b). $\lambda_{exc}$ = 514 nm. (c) are bright field images and (d) are corresponding overlay images. Brightness and contrast levels in the bright field channel have been adjusted for better visualizing the fibers. Scale bars: 20 µm. No fluorescence is detected from PLLA fibers under the different pHs.

## References

[1]    L. L. del Mercato, A. Z. Abbasi, W. J. Parak, *Small* **2011**, 7, 351.

[2]    M. De Luca, M. M. Ferraro, R. Hartmann, P. Rivera-Gil, A. Klingl, M. Nazarenus, A. Ramirez, W. J. Parak, C. Bucci, R. Rinaldi, L. L. del Mercato, *ACS Appl. Mater. Interfaces* **2015**, 15, 15052.